\documentclass[conference]{IEEEtran}
\IEEEoverridecommandlockouts
\usepackage{cite}
\usepackage{amsmath,amssymb,amsfonts}
\usepackage{algorithmic}
\usepackage{graphicx}
\usepackage{textcomp}
\usepackage{xcolor}
\usepackage{mathtools}
\usepackage{commath}
\usepackage{hyperref}
\usepackage{hyperref}
\hypersetup{colorlinks,allcolors=black}
\usepackage[compact]{titlesec}
\titlespacing{\section}{0pt}{2ex}{1ex}
\titlespacing{\subsection}{0pt}{1ex}{0ex}
\titlespacing{\subsubsection}{0pt}{0.5ex}{0ex}

\def\BibTeX{{\rm B\kern-.05em{\sc i\kern-.025em b}\kern-.08em
    T\kern-.1667em\lower.7ex\hbox{E}\kern-.125emX}}
    
\begin{document}
 
\title{Deep Learning Phase Compression for MIMO CSI Feedback 
by Exploiting FDD Channel Reciprocity}

\author{Yu-Chien Lin, Zhenyu Liu, Ta-Sung Lee, and Zhi Ding
\thanks{Y.-C Lin is with the Department of Electrical and Computer Engineering,
University of California at Davis, Davis, CA, USA, and
was affiliated with National Yang Ming Chiao Tung University, Taiwan (e-mail: ycmlin@ucdavis.edu).

Z. Liu was with University of California, Davis, CA, USA and is now with Beijing University of Posts and Telecommunications, China (e-mail: lzyu@bupt.edu.cn).

T.-S Lee is with the Institute of Communications Engineering, National Yang Ming Chiao Tung University, Taiwan (e-mail: tslee@mail.nctu.edu.tw).

Z. Ding is with the Department of Electrical and Computer Engineering,
University of California, Davis, CA, USA (e-mail: zding@ucdavis.edu).}

\thanks{This work was supported by the Center for Open Intelligent Connectivity from The Featured Areas Research Center Program within the framework of the Higher Education Sprout Project by the Ministry of Education of Taiwan.}}

\maketitle

\begin{abstract}

Large scale MIMO FDD systems are often hampered 
by bandwidth required to feedback
downlink CSI. 
Previous works have made notable progresses  
in efficient CSI encoding and recovery
by taking advantage of FDD uplink/downlink 
reciprocity between their CSI magnitudes.  
Such framework separately encodes CSI 
phase and magnitude. To further enhance feedback efficiency, 
we propose a new deep learning architecture 
for phase encoding based on limited 
CSI feedback and magnitude-aided information. 
Our contribution features a framework with a modified loss function 
to enable end-to-end joint optimization of CSI magnitude 
and phase recovery. 
Our test results show superior 
performance in indoor/outdoor scenarios.
\end{abstract}


\section{Introduction}

Massive multiple-input multiple-output (MIMO) transceiver systems have demonstrated significant success in achieving high spectrum and energy efficiency for 5G and future wireless communication systems. 
These high achievable benefits require sufficiently accurate 
downlink (DL) channel state information (CSI) at the gNB (i.e., gNodeB). 
Frequency-division duplexing (FDD) systems, however, 
can only estimate DL CSI through feedback from UEs because DL and uplink (UL) channels occupy different frequency bands and may exhibit different
channel characteristics. 
Since the feedback resource overhead grows 
proportionally with increasing MIMO size
and spectrum, reducing CSI feedback overhead 
is vital to the widespread deployment of
FDD MIMO systems. To improve feedback efficiency, previous works in \cite{CsiNet, CsiNet+} 
developed a deep neural network with an autoencoder structure 
whose encoder at UEs and base stations, respectively
for CSI compression and recovery. Related
works and variants \cite{CsiNetVar3} have  
demonstrated performance advantages over
traditional compressive sensing approaches.

Recent works have revealed the importance of exploiting
correlated channel information such as
UL CSI \cite{DualNet, CQNET}, past CSI \cite{MarkovNet}, and CSI adjacent UEs \cite{CoCsiNet} for improving the
accuracy of DL CSI recovery at base stations. 
In particular, important physical insights 
regarding FDD reciprocity, slow changes in propagation 
environment in time, and similar 
propagation conditions within short geographical 
distance, underscore respectively the strong spectral, temporal, and spatial correlations between
magnitudes of different CSI in delay-angle (DA) domain.
Since, side information from correlated CSI 
lowers the conditional entropy (uncertainty) of the 
DL CSI, their effective utilization
reduces encoded feedback payload
required from UEs \cite{MarkovNet}. 

Unlike the DA domain CSI magnitudes which tend to clearly exhibit strong temporal, spectral, and spatial correlations, CSI phases are very sensitive to
changes in time, frequency, and location. 
To exploit the side CSI information, 
existing solutions have adopted dual feedback framework which separately encodes and recovers CSI phases
from their corresponding magnitudes. These studies have utilized an isolated autoencoder to compress and recover the CSI magnitudes. 
For phase recovery, the basic principle in \cite{DualNet, CQNET, CoCsiNet} is to expend more feedback resources (bandwidth) to encode the significant phases 
according to the corresponding magnitudes. 
For example, the authors in \cite{CoCsiNet} designed a deep learning model with a magnitude-dependent polar-phase (MDPP) loss function to 
compress the significant CSI phases depending on the CSI magnitude. 

Presently, these existing magnitude-dependent
CSI feedback frameworks tend to train two learning
models to encode and recover 
DL CSI magnitudes and phases, respectively. 
Intuitively, both CSI magnitudes and phases depend on
the RF propagation environment including multipath
delays, Doppler spread, bandwidth, and scatter distribution. 
CSI magnitude and phase encoding and recovery
should be jointly instead of individually optimized. 
In fact, the structural sparsity of CSI phases
and their joint distribution with the CSI magnitude
are generally unknown and under explored.

In this work, we develop a deep learning based 
CSI feedback framework which jointly optimizes the
magnitude and phase encoding. We propose
a new loss function, namely
sinusoidal magnitude-adjust phase error (SMAPE), 
that directly corresponds to the MSE of DL CSI recovery. 
Furthermore, we take advantage of 
the circular properties of
CSI matrices in DA domain and propose
a novel circularly convolutional neural network (C-CNN) 
layers proved to significantly enhance 
CSI compression efficiency and recovery performance.

\section{System Model}

Without loss of generality, we
consider a single-cell MIMO FDD link in which
a gNB with $N_b$ antennas communicates with a single
antenna UE. 
The OFDM signal spans $N_f$ DL subcarriers.
The DL received signal of the $k$th subcarrier is 
\begin{equation}
\setlength{\abovedisplayskip}{4pt}
\setlength{\belowdisplayskip}{4pt}
y_\text{DL}^{(k)}= \mathbf{h}_\text{DL}^{(k)H} \mathbf{w}_\text{T}^{(k)}x_\text{DL}^{(k)} + n_\text{DL}^{(k)}\label{DL signal},
\end{equation}
where $(\cdot)^H$ denotes  conjugate transpose. 
Here for the $k-$th subcarrier, $\mathbf{h}_\text{DL}^{(k)}\in \mathbb{C}^{N_b \times 1}$ denotes the
CSI vector and $\mathbf{w}_\text{T}^{(k)}\in \mathbb{C}^{N_b \times 1}$ denotes the corresponding precoding vector\footnote{gNB calculates precoding vectors at subcarriers with DL CSI matrix.} whereas
$x_\text{DL}^{(k)}\in \mathbb{C}$ and $n_\text{DL}^{(k)}\in \mathbb{C}$ denote the DL source signal and additive
noise, respectively. With the same antennas, gNb receives 
UL signal
\begin{equation}
\setlength{\abovedisplayskip}{4pt}
\setlength{\belowdisplayskip}{4pt}
\mathbf{y}_\text{UL}^{(k)}= \mathbf{h}_\text{UL}^{(k)}x_\text{UL}^{(k)} + \mathbf{n}_\text{UL}^{(k)}\in \mathbb{C}^{N_b \times 1}\label{UL signal},
\end{equation}
{where subscript UL is 
used to
denote uplink channel
$\mathbf{h}_\text{UL}^{(k)}\in \mathbb{C}^{N_b \times 1}$,
signal $x_\text{UL}^{(k)}$, and noise
$\mathbf{n}_\text{UL}^{(k)}\in \mathbb{C}^{N_b \times 1}$. 
DL and UL channel vectors can be written
as spatial-frequency CSI (SF-CSI) matrices ${\mathbf{H}}_{\text{DL}}^\text{SF} = [\mathbf{h}_\text{DL}^{(1)},...,\mathbf{h}_\text{DL}^{(N_f)}]^H \in \mathbb{C}^{N_f \times N_b}$ and ${\mathbf{H}}_{\text{UL}}^\text{SF} = [\mathbf{h}_\text{UL}^{(1)},...,\mathbf{h}_\text{UL}^{(N_f)}]^H \in \mathbb{C}^{N_f \times N_b}$, respectively.
Typically in FDD systems, DL CSI  ${\mathbf{H}}_{\text{DL}}^\text{SF}$ is estimated by UE for feedback to gNB. However, the
number ($N_f\times N_b$) of unknowns in  ${\mathbf{H}}_{\text{DL}}^\text{SF}$ occupies
substantial feedback spectrum in large or massive
MIMO systems. To exploit CSI sparsity
to reduce feedback overhead,
we apply  IDFT
$\mathbf{F}_{D}\in \mathbb{C}^{N_f \times N_f}$ and DFT $\mathbf{F}_{A}\in \mathbb{C}^{N_b \times N_b}$ on
$\mathbf{H}^{\text{SF}}$ to generate 
DA domain CSI matrix}
\begin{equation}
\setlength{\abovedisplayskip}{4pt}
\setlength{\belowdisplayskip}{4pt}
\mathbf{H}^{\text{DA}} = \mathbf{F}_{D}\mathbf{H}^{\text{SF}}\mathbf{F}_{A},\label{DL CSI matrix in angle-delay domain}
\end{equation}
which demonstrates sparsity. Note that $\mathbf{H}^\text{SF}$ denotes either $\mathbf{H}^\text{SF}_\text{UL}$ or $\mathbf{H}^\text{SF}_\text{UL}$. Owing to limited multipath
delay spread and limited number of scatters,
most elements in $\mathbf{H}^{\text{DA}}$ 
are found to be near insignificant, 
except for the first $Q_f$ and the last $Q_l$ rows.
Therefore, we shorten the CSI matrix $\mathbf{H}^{\text{DA}}$ in DA domain
to $Q_t = Q_f + Q_l$ rows that contain 
sizable non-zero values and 
utilize $\mathbf{H}_\text{DL}$ and $\mathbf{H}_\text{UL}$ to denote the correspondingly truncated matrices of
${\mathbf{H}}_\text{DL}^\text{DA}$ and ${\mathbf{H}}_\text{UL}^\text{DA}$, respectively. 
For simplicity, we shall denote $\mathbf{H}_\text{DL}$ as $\mathbf{H}$ in the rest of this work except for
cases when ambiguity may arise.  

Subsequently, to further reduce the feedback overhead, the DL CSI matrix $\mathbf{H}$ is encoded 
at the UE and recovered by the gNB. The recovered DL CSI matrix can be expressed as
\begin{equation}
\setlength{\abovedisplayskip}{3pt}
\setlength{\belowdisplayskip}{3pt}
 \widehat{\mathbf{H}} = f_{\text{de}}(f_{\text{en}}(\mathbf{H})),\label{Recovered DL CSI matrix}
\end{equation}
where $f_{\text{en}}(\cdot)$ and $f_{\text{de}}(\cdot)$ denote encoding/decoding operations.

\section{Magnitude-aided CSI feedback Framework}
\begin{figure}[tb]
\setlength{\abovecaptionskip}{0.cm}
\setlength{\belowcaptionskip}{-0.cm}
\centering
\resizebox{3.2in}{!}{
\includegraphics*{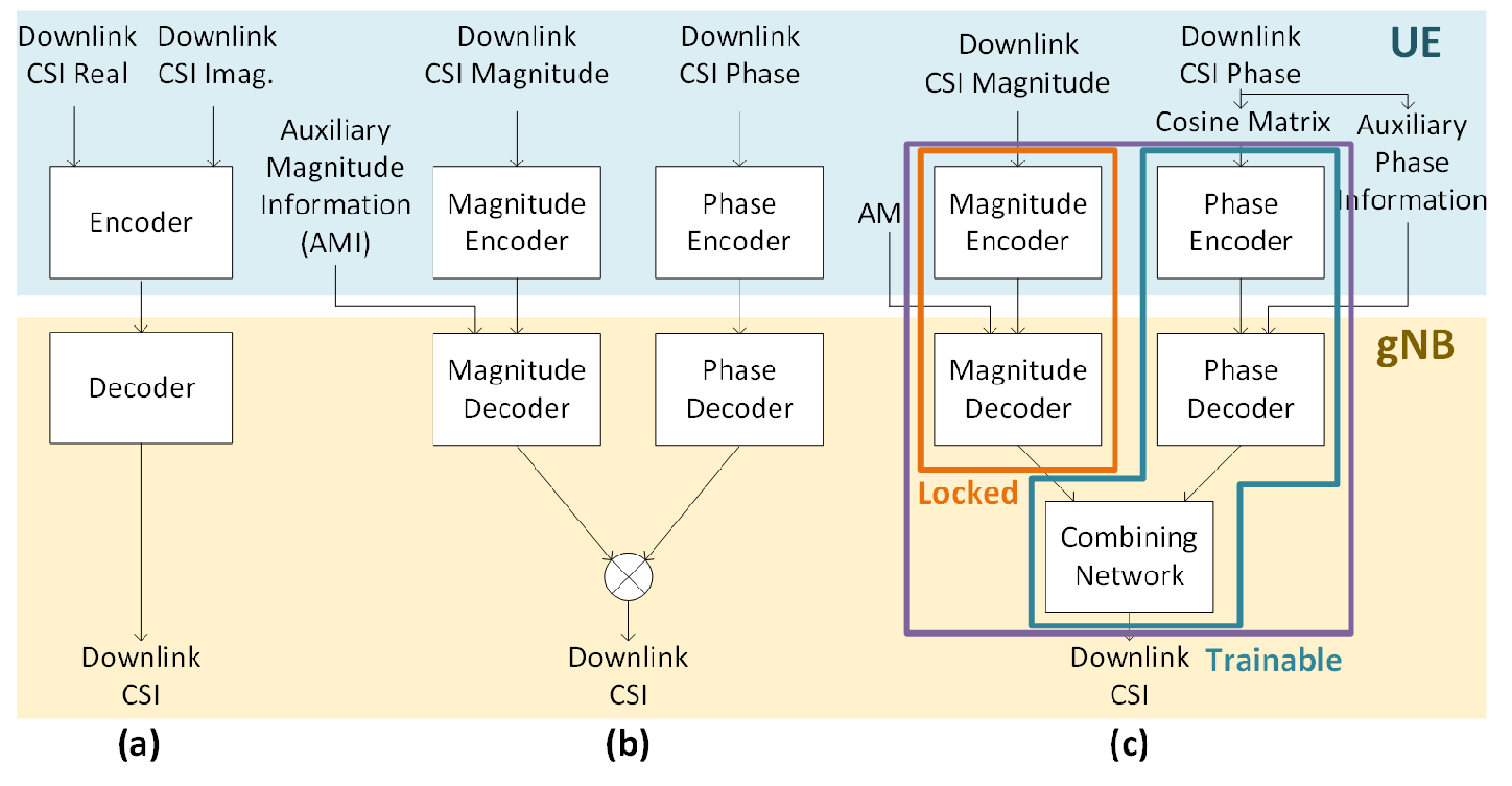}}
\caption{General network architecture. (a) Conventional CSI feedback framework, (b) conventional magnitude-aided CSI feedback framework, and  (c) proposed magnitude-aided CSI feedback framework.\label{GA}}
\end{figure}

Most deep-learning works on CSI 
compression leverage the success of real-valued
deep learning network (DLN) in image processing by separating CSI matrices
into real and imaginary parts that are analogous
to image files
\cite{CsiNet, CsiNet+, MarkovNet}, as shown in 
Fig.~\ref{GA}(a).  
Recent studies \cite{DualNet, MarkovNet, CoCsiNet}, 
however, uncovered the benefit of 
separately encoding  magnitudes and
phases of $\mathbf{H}$ instead in order to better
exploit other correlated CSI magnitudes as 
auxiliary magnitude information (AMI).
Such architecture,
illustrated in Fig.~\ref{GA}(b), requires
substantially lower feedback overhead for the 
magnitudes of $\mathbf{H}$
and allocate more feedback resources
for phase feedback of $\mathbf{H}$.

Fig.~\ref{GA}(c) illustrates
our proposed new DLN framework,
consisting of magnitude and phase branches.
The gNB further contains a combining network 
to estimate the full CSI based on
results from magnitude and phase decoders.
We optimize encoders, decoders, and
combining network jointly 
by minimizing a single loss function 
for end-to-end learning during offline training. 
Note that the magnitude branch can be 
independently optimized. {To
facilitate convergence \cite{preprint}, the training of the DLN 
has two stages.} In stage-1, the CSI magnitude 
encoder/decoder branch is pre-trained for 
magnitude recovery. In stage-2, 
both the CSI phase branch and the combining network are optimized with the help of the magnitude branch, while the parameters of the magnitude branch are fixed. 

\subsection{DualNet-MP}
\begin{figure}
\setlength{\abovecaptionskip}{0.cm}
\setlength{\belowcaptionskip}{10cm}
\centering
\resizebox{3.5in}{!}{
\includegraphics*{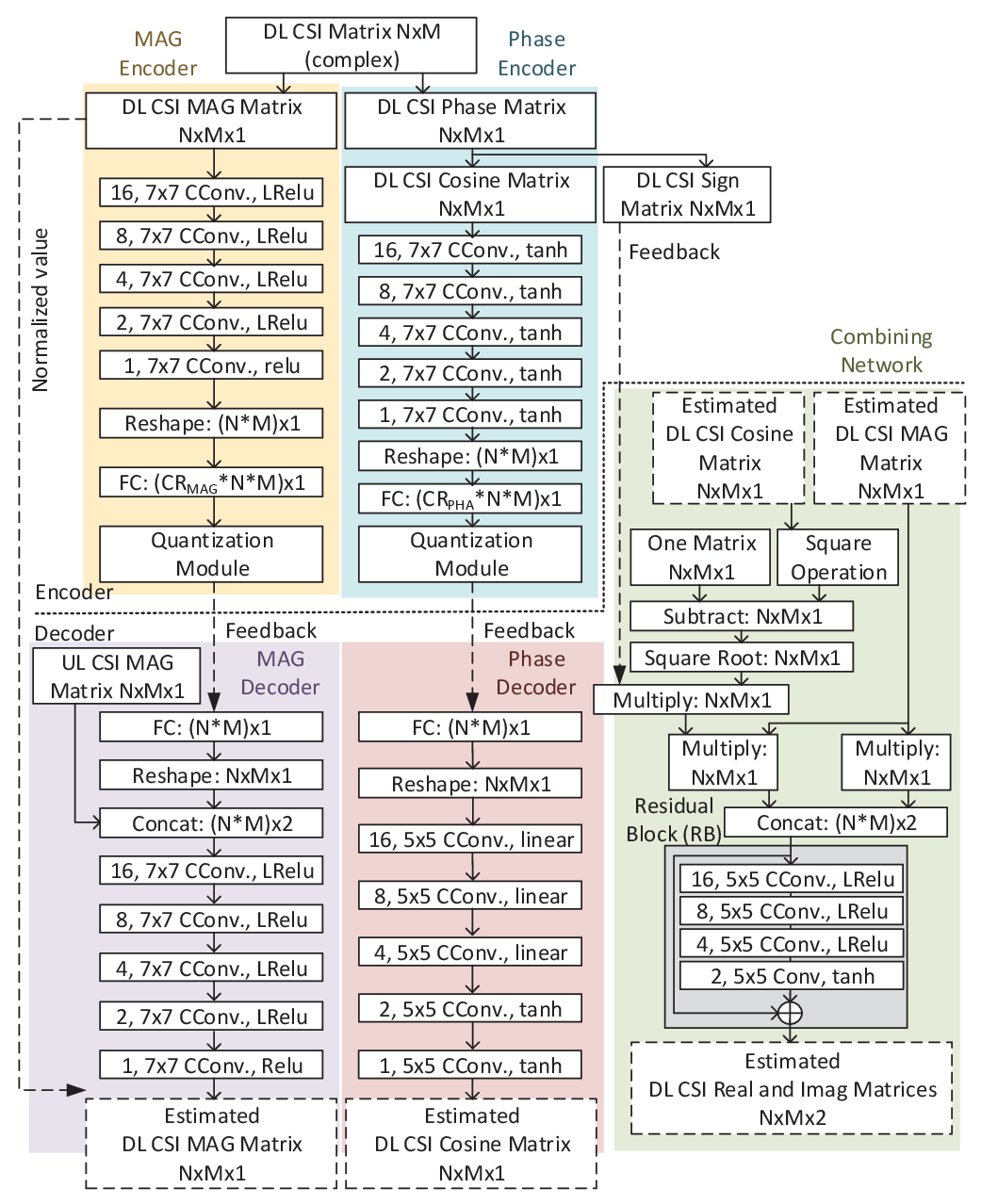}}
\caption{Network architecture of DualNet-MAG-PHA.\label{DualNet-MAG-PHA}}
\end{figure}

We now present a new DLN called DualNet-MP. 
As shown in Fig. \ref{DualNet-MAG-PHA}, DualNet-MP splits each complex CSI matrix 
into
\[\mathbf{H}=|\mathbf{H}| \odot e^{j\measuredangle{\mathbf{H}}},
\]
where $\odot$ represents Hadamard product.
Denote the $(m,n)$-th entry of $\mathbf{H}$ as
$\mathbf{H}_{m,n}=|\mathbf{H}_{m,n}|e^{j\measuredangle{\mathbf{H}_{m,n}}}$.
The magnitude matrix 
$|\mathbf{H}|$ and consists of entries $|\mathbf{H}_{m,n}|$ and 
phase matrix $e^{j\measuredangle{\mathbf{H}}}$ consists of
entries $e^{j\measuredangle{\mathbf{H}_{m,n}}}$.

Similar to \cite{CQNET}, we forward the CSI magnitudes 
to the magnitude encoder network, including four $7 \times 7$ circular convolutional layers with 16, 8, 4, 2 and 1 channels and activation functions. Given the circular characteristic of CSI matrices, we introduce circular convolutional layers to replace the traditional linear ones.
{Subsequently, a fully connected (FC) layer with $\left\lceil{\text{CR}_\text{MAG}Q_tN_b}\right\rceil$ elements is included for dimension reduction after reshaping.} $\text{CR}_\text{MAG}$ denotes the magnitude compression ratio. The output of the FC layer is then fed into the quantization module, called the sum-of-sigmoid (SSQ) \cite{CQNET} to generate magnitude codewords for feedback.

{At the gNB, a magnitude decoder uses magnitude codewords and available UL CSI magnitudes\footnote{ UL CSI is estimated at the gNB and assumed to be perfectly estimated.} as AMI to jointly 
recover DL CSI magnitudes. The DL magnitude branch is first optimized by updating the network parameters $\Theta_{\text{MAG}}$}
\begin{equation}
\setlength{\abovedisplayskip}{4pt}
\setlength{\belowdisplayskip}{4pt}
{\mathop{\arg\min}_{\Theta_\text{en,MAG},\Theta_\text{de,MAG}} \left\{
\norm{|\mathbf{\hat{H}}|-|\mathbf{H}|}^2_\text{F}
\right\}} \\ \label{loss_magnitude}
\end{equation}
{to minimize the MSE of recovered MIMO CSI magnitude}
\begin{equation}
{|\mathbf{\hat{H}}| = f_\text{de,MAG}(f_\text{en,MAG}(|\mathbf{H}|,\Theta_{\text{en,MAG}}),\Theta_\text{de,MAG},\mathbf{H}_\text{UL}),}
\end{equation}
{in which $\Theta$ denotes DLN parameters and subscripts en, de, UL, and MAG of the $f(\cdot)$  denote the encoder, decoder, UL, and magnitude branch, respectively.}
    
For CSI recovery of MIMO channels, we are only 
interested in their
wrapped phases (i.e., $\measuredangle{\mathbf{H}}$).
There is 1-to-1 relationship between a phase value
$\phi$ and $\left(\cos(\phi), \mbox{sign}[\sin(\phi)]\right)$.
For these reasons, we propose to form a "cosine"
matrix whose entries are cosines of entries from $\mathbf{H}$
denoted by
\begin{subequations}
\begin{equation}
\setlength{\abovedisplayskip}{3pt}
\setlength{\belowdisplayskip}{3pt}
\mathbf{Cos}=\cos (\measuredangle{\mathbf{H}}).
\end{equation}
{Let $\mathbf{A}_{m,n}=\mbox{sign} [\sin[\measuredangle(\mathbf{H}_{m,n})]]$.
We also form a sign matrix} 
\begin{equation}
\setlength{\abovedisplayskip}{3pt}
\setlength{\belowdisplayskip}{3pt}
{\mathbf{A} = \left[\mathbf{A}_{m,n}\right].}
\end{equation}
\end{subequations}
{Thus 
$(\mathbf{Cos}, \mathbf{A})$ uniquely determines
$\measuredangle{\mathbf{H}}$. }

Since $\mathbf{Cos}$ matrix is real, we can
adopt a phase encoder similar to the magnitude encoder.
Let $\text{CR}_\text{PHA}$ denotes the phase compression
ratio. Each $\mathbf{Cos}$ generates a
$\lceil\text{CR}_\text{PHA}Q_tN_b\rceil$-element codeword.
Our DLN uses tanh activation function 
in each circular convolutional layer of the phase encoder 
to capture the underlying features of significant phases associated with large magnitudes.
Upon completion of encoder training, 
the UE processes each CSI $\mathbf{H}$, and feeds back
the CSI magnitude codeword $f_\text{en,MAG}(|\mathbf{H}|,\Theta_{\text{en,MAG}})$, the phase codeword $f_\text{en,PHA}(\mathbf{Cos},\Theta_{\text{en,PHA}})$ 
and the sign matrix $\mathbf{A}$ to gNB. 

At the gNB receiver, the phase codeword $f_\text{en,PHA}(\mathbf{Cos},\Theta_{\text{en,PHA}})$ and the feedback sign matrix $\mathbf{A}$ are sent
to the phase decoder with the tanh activation function as the last layer to constrain the entries of DL CSI cosine matrix $\widehat{\mathbf{Cos}}$ within $[-1, 1]$.
The magnitude codeword and side information are used
by the magnitude decoder to obtain an 
estimated CSI magnitude matrix $|\widehat{\mathbf{H}}|$.
Based on the relationship
$\sin(\phi) =\mbox{sign}(\sin[\phi]) \sqrt{1- \cos^2 (\phi)}$,
we form
\[
\widehat{\mathbf{Sin}}=\mathbf{A}\odot(\mathbf{1}-\widehat{\mathbf{Cos}}\odot\widehat{\mathbf{Cos}})^{1/2}.\]
Therefore, we can directly generate a preliminary
CSI estimate
\[\widehat{\mathbf{H}}=\left[|\widehat{\mathbf{H}}|
\odot \widehat{\mathbf{Cos}}, 
|\widehat{\mathbf{H}}|\odot \widehat{\mathbf{Sin}}\right]
\] 
from locally available
$\widehat{\mathbf{Cos}}$, $\mathbf{A}$, 
$\widehat{|\mathbf{H}|}$.
The combining network is trainable and can include two residual blocks containing four circular convolutional layers to refine the DL CSI matrix. 

{For end-to-end optimization,
our training criterion relies on}
\begin{equation}
\setlength{\abovedisplayskip}{3pt}
\setlength{\belowdisplayskip}{3pt}
{\mathop{\text{minimize}}_{\Theta_\text{en,PHA},\Theta_\text{de,PHA},\Theta_\text{C}} \left\{
\norm{\mathbf{\hat{H}}-\mathbf{H}}^2_\text{F}
\right\},} \\ \label{loss_CSIMSE}
\end{equation}
{to optimize the parameters 
$  \Theta_\text{en,PHA}$ of phase encoder $f_\text{en,PHA}$ and
parameters 
$ \Theta_\text{de,PHA}$ of phase decoder 
$f_\text{en,PHA}$ to estimate}
\begin{equation}
\setlength{\abovedisplayskip}{4pt}
\setlength{\belowdisplayskip}{4pt}
\widehat{\mathbf{Cos}} = f_\text{de,PHA}(f_\text{en,PHA}(\mathbf{Cos},\Theta_{\text{en,PHA}}), \mathbf{A}, \Theta_\text{de,PHA}).
\end{equation}
Using the same loss function
(\ref{loss_CSIMSE}),
we also train the combining
network $f_\text{C}$ by optimizing
parameters $\Theta_\text{C}$ to generate
\begin{equation}
\mathbf{\hat{H}} = f_\text{C}(|\widehat{\mathbf{H}}|,\widehat{\mathbf{Cos}},\Theta_\text{C}).
\end{equation}
Since the
training of the magnitude learning 
branch can be decoupled, 
our framework optimizes the entire 
architecture by minimizing the overall CSI MSE
of (\ref{loss_CSIMSE}). It is possible, however,
to also partially incorporate the 
MSE of (\ref{loss_CSIMSE}) 
to further refine the magnitude DLN branch
by adopting a slower learning rate. 
 
\subsection{Loss Function Redesign}

Considering the MSE loss function, 
it may be intuitive to simply rewrite the loss function 
as follows: 
\begin{equation}\label{loss_MSE_direct}
\begin{aligned}
&Loss_\text{0} = \text{MSE}_\text{CSI}(\widehat{|\mathbf{H}|},
\measuredangle{\hat{\mathbf{H}}}) = 
\norm{\mathbf{H} - \mathbf{\hat{H}}}^2_\text{F}\\
& = \norm{|\mathbf{H}|\odot\text{cos}(\measuredangle{\mathbf{H}}) -|\hat{\mathbf{H}}|\odot\text{cos}(\measuredangle{\hat{\mathbf{H}}})}^2_\text{F}  \\ 
& + \norm{|\mathbf{H}|\odot\text{sin}(\measuredangle{\mathbf{H}}) -|\hat{\mathbf{H}}|\odot\text{sin}(\measuredangle{\hat{\mathbf{H}}})}^2_\text{F}. 
\\
\end{aligned}
\end{equation}
This means that $|\mathbf{H}|$ and 
$\measuredangle{\mathbf{H}}$ are used as
encoder network input variables whereas their estimates
are the decoder network output variables. However, the presence of infinitely many and shallow local minima of sinusoidal functions 
$\sin(\cdot)$ and $\cos(\cdot)$ often lead to training difficulties \cite{TTW}.  To overcome this problem, the authors in \cite{CoCsiNet} recently proposed a weighted 
MDPP loss function 
\begin{equation}
\text{MSE}_\text{MDPP} = 
\norm{|\measuredangle{\mathbf{H}} - \measuredangle{\mathbf{\hat{H}}}|{\odot}|\mathbf{H}|}^2_\text{F}  \\ \label{loss_magnitude_dependent_phase}
\end{equation}
which still uses $|\mathbf{H}|$ and 
$\measuredangle{\mathbf{H}}$ as input and output 
variables. 
where $\measuredangle{\mathbf{H}}$ and $\measuredangle{\mathbf{\hat{H}}}$ denote the true and estimated phases, respectively. By weighting 
the original phase discrepancy 
with the true CSI magnitude,
this new loss function helps capture the underlying features of the critical phases associated with CSI coefficents with
dominant magnitudes.
However, the loss function is not equivalent to our final goal
for minimizing MSE of DL CSI.
We now propose a reparamterization of the same MSE loss function during training.  Instead of changing the loss function, 
we can overcome the training problem of
directly parameterization in
Eq. (\ref{loss_MSE_direct}). Instead, recognizing that
only the wrapped phases of $\measuredangle{\mathbf{H}}$ are
of interest, we replace
$\measuredangle{\mathbf{H}}$ with $\mathbf{Cos}$ 
and $\mathbf{A}$ via
the following reparameterization: 
\begin{align}
\label{loss_proposed}
\text{MSE}_\text{SMAPE} (\widehat{|\mathbf{H}|},
\widehat{\mathbf{Cos}},\mathbf{A})&= 
\norm{\mathbf{H} - \mathbf{\hat{H}}}^2_\text{F} \\
& = \norm{|\mathbf{H}|\odot\mathbf{Cos} -|\hat{\mathbf{H}}|\odot\widehat{\mathbf{Cos}}}^2_\text{F}  
\nonumber \\ 
& + \norm{|\mathbf{H}|\odot\mathbf{Sin} -|\hat{\mathbf{H}}|\odot\widehat{\mathbf{Sin}}}^2_\text{F},
\nonumber
\end{align}
where we have used the sign matrix 
$\mathbf{A}$ feedback to generate
\begin{subequations}
\begin{align}\label{loss_proposed_where2}
{\mathbf{Sin}} &= \mathbf{A}\odot(\mathbf{1}-{\mathbf{Cos}}\odot{\mathbf{Cos}})^{1/2}\\
\widehat{\mathbf{Sin}} &= \mathbf{A}\odot(\mathbf{1}-\widehat{\mathbf{Cos}}\odot\widehat{\mathbf{Cos}})^{1/2}.
\end{align}
\end{subequations}
This formulation saves about half the bandwidth by 
sending the sign matrix $\mathbf{A}$ without
encoding matrix $\mathbf{Sin}$.

Moreover, the sparsity of $\mathbf{H}$ means
that we only need to
feed back partial entries of $\mathbf{A}$ associated with 
a swath of entries with dominant magnitudes.
If we define a reduction ratio $R_\text{s}$ to further reduce feedback overhead\footnote{Usually, the reconstruction performance can remain approximately the same even if the sign ratio $R_\text{s}$ is less than $0.25$ due to the sparsity.}. The total phase feedback overhead (in bits) is summarized as follows: 
\begin{equation}
    B_\text{SMAPE} = \text{CR}_\text{PHA}(K_\text{PHA}Q_{t}N_{b} + R_\text{s}Q_{t}N_{b}) (\text{bits}),
\end{equation}
where $K_\text{PHA}$ denotes the number of encoding
bits for each entry of the compressed cosine matrix $f_\text{en,PHA}(\mathbf{Cos},\Theta_{\text{en,PHA}})$. 

To summarize our training strategy of DualNet-MP, 
we use Eq. (\ref{loss_magnitude}) 
as the loss function during the first training stage. 
In the second training stage, we used Eqs. (\ref{loss_proposed}) as the loss function to build an end-to-end learning architecture.

\section{Experimental Evaluations}

\subsection{Experiment Setup}
In our experiments\footnote{The source codes are online available in \href{https://github.com/max821002/DualNet-MP/}{https://github.com/max821002/DualNet-MP/}}, we let the UL and DL bandwidths be 20 MHz and the subcarrier number be $N_f$ = 1024.
We consider both indoor and outdoor cases. 
We place the gNB with a height of 20 m at the center of 
a circular cell coverage
with a radius of 20 m for indoor and 200 m for outdoor. 
The number of gNB antennas is $N_b = 32$ whereas each UE 
has a single antenna. A half-wavelength inter-antenna spacing is considered. 
For each trained model, the number of epochs and batch size were set to 1,000 and 200, respectively. {We generate two datasets each containing 100,000 random channels for both indoor and outdoor cases from two different channel models. 
60,000 and 20,000 random channels are for training and validation. 
The remaining 20,000 random channels are test data for performance evaluation.}

In the first (indoor) dataset, we used the COST 2100 \cite{COST2100} simulator and select the scenario \textit{IndoorHall at 5GHz} to generate indoor channels at 5.1-GHz UL and 5.3-GHz DL with LOS paths. The antenna and band types are set as \textit{MIMO VLA omni} and \textit{wideband}, respectively. As for the outdoor dataset, we utilized QuaDRiGa simulator \cite{QuaDriGa} with the scenario features of \textit{3GPP 38.901 UMi}.
We considered the UMi scenario at 2 and 2.1 GHz of UL and DL bands, respectively, without LOS paths. The number of cluster paths was set as 13. The antenna type is set to \textit{omni}. 
By DFT/IDFT operations and truncating described in Eqs. (3) and (4) of the revised manuscript, the resulting $32\times32$ UL/DL CSI matrices in DA domain are obtained and be fed as the encoder input after proper processes.

The performance metric is the normalized MSE
\begin{equation}
\text{NMSE} =  \frac{1}{D}\sum^{D}_{d=1}\norm{\widehat{\mathbf{H}}_{\text{DL},d}^\text{SF} - {\mathbf{H}}_{\text{DL},d}^\text{SF}}^2_\text{F} /\norm{{\mathbf{H}}_{\text{DL},d}^\text{SF}}^2_\text{F},\label{NMSE1}
\end{equation}
where the number $D$ and subscript $d$ denote the total number and index of channel realizations, respectively. Instead of evaluating the estimated DL CSI matrix $\widehat{\mathbf{H}}_\text{DL}$, we evaluate the estimated SFCSI matrix $\widehat{\mathbf{H}}_\text{DL}^\text{SF}$ that can be obtained by reversing the Fourier processing and padding zero matrix.
Note that this NMSE includes both the
errors caused by truncation at the encoder
and the overall recovery error. Thus, it
is practically more meaningful. 

In the following section, we evaluate the performance 
of CSI recovery by adopting the proposed optimization
method
and encoder/decoder architecture. Thus, we trained DualNet-MP 
with the same core network design for magnitude recovery.
However, we test different methods to reconstruct the CSI phases 
for two phase compression ratios 
of $\text{CR}_\text{PHA} = 1/8$ and $1/16$
\footnote{All alternate approaches consume $1.2$ and $0.625$ bits/phase entry}:
\begin{itemize}
\item SMAPE: the network architecture follows DualNet-MP as illustrated in Fig. 2 of the revised manuscript. The sign ratio $R_\text{s}$ varies
between $[0.25,0.125]$ and we use
$K_\text{PHA} = 8$ bits for both $\text{CR}_\text{PHA} = \{1/8,1/16\}$. The loss function for phase reconstruction is given as follows:
\begin{equation}
\begin{aligned}
\label{loss_proposed}
\text{MSE}_\text{SMAPE} (\widehat{|\mathbf{H}|},
\widehat{\mathbf{Cos}},\mathbf{A})&= 
\norm{\mathbf{H} - \mathbf{\hat{H}}}^2_\text{F} \\
& = \norm{|\mathbf{H}|\odot\mathbf{Cos} -|\hat{\mathbf{H}}|\odot\widehat{\mathbf{Cos}}}^2_\text{F}  
\nonumber \\ 
& + \norm{|\mathbf{H}|\odot\mathbf{Sin} -|\hat{\mathbf{H}}|\odot\widehat{\mathbf{Sin}}}^2_\text{F}.
\nonumber
\end{aligned}
\end{equation}
\item MDPQ \cite{DualNet}: the design assigns more encoding bits to encode the significant phases according the corresponding magnitudes. It assigns $[0, 0, 0, 3, 7]$ and $[0, 0, 0, 0, 5]$ bits for $\text{CR}_\text{PHA} = [1/8,1/16]$, respectively, to encode the CSI phases corresponding to $[0, 0.5, 0.7, 0.8, 0.9]$ of the cumulative distribution of CSI magnitude. 
\item MSE${}_0$: instead of cosine, CSI phases are fed directly
to the phase encoder. 
Both cosine and sine functions are appended as the final
layer of the phase decoder to transform the decoded phase into cosine and sine, respectively,
to recover the real and imaginary parts of CSI. We set $K_\text{PHA}$ to $8$ bits. The loss function for phase reconstruction is given as follows:
\begin{equation}\label{loss_MSE_direct}
\begin{aligned}
&\mbox{Loss}_\text{0} = \text{MSE}_\text{CSI}(\widehat{|\mathbf{H}|},
\measuredangle{\hat{\mathbf{H}}}) = 
\norm{\mathbf{H} - \mathbf{\hat{H}}}^2_\text{F}\\
& = \norm{|\mathbf{H}|\odot\text{cos}(\measuredangle{\mathbf{H}}) -|\hat{\mathbf{H}}|\odot\text{cos}(\measuredangle{\hat{\mathbf{H}}})}^2_\text{F}  \\ 
& + \norm{|\mathbf{H}|\odot\text{sin}(\measuredangle{\mathbf{H}}) -|\hat{\mathbf{H}}|\odot\text{sin}(\measuredangle{\hat{\mathbf{H}}})}^2_\text{F}. 
\\
\end{aligned}
\end{equation}

\item MDPP \cite{CoCsiNet}: we reuse the loss function Eq.(\ref{loss_magnitude_dependent_phase}) with the same network architecture. We set $K_\text{PHA}$
to $10$ bits.

\begin{equation}
\text{MSE}_\text{MDPP} = 
\norm{|\measuredangle{\mathbf{H}} - \measuredangle{\mathbf{\hat{H}}}|{\odot}|\mathbf{H}|}^2_\text{F}  \\ \label{loss_magnitude_dependent_phase}
\end{equation}

\end{itemize}
\subsection{Different Phase Compression Designs}
\begin{figure}
    \setlength{\abovecaptionskip}{0.cm}
    \setlength{\belowcaptionskip}{-1cm}
\centering
\resizebox{3.8in}{!}{
\includegraphics*{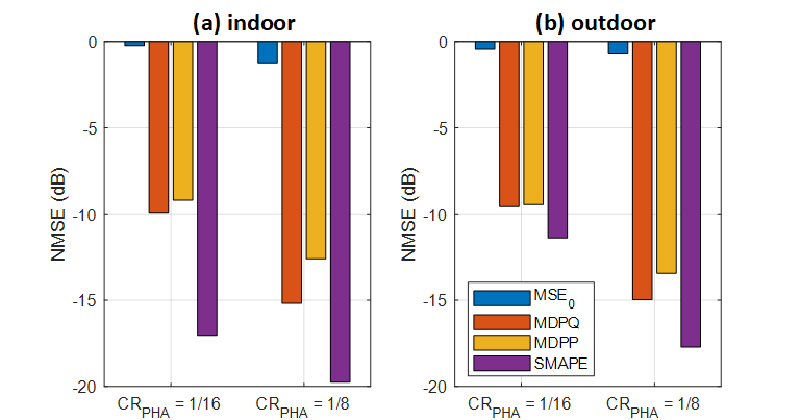}}
\caption{NMSE performance for different loss functions in (a) indoor and (b) outdoor scenarios.\label{exp_B}}
\end{figure}

To demonstrate the superiority of the proposed SMAPE loss
function, we applied different phase reconstruction approaches to 
DualNet-MP
for different phase compression ratios $\text{CR}_\text{PHA}$. 
Figs. \ref{exp_B} (a) and (b) show the NMSE performance of
different approaches under indoor and outdoor scenarios,
respectively, at different compression ratios. As expected, DaulNet-MP
encounters training difficulties when using the
simple loss function $Loss_0$. 
By adopting MDPP loss functions, DualNet-MP performs 
much better than the simple loss function $Loss_0$. Although DualNet-MP appears to be better when using MDPQ instead of MDPP,
encoding bit-assignment require careful tuning to achieve a satisfactory result. Finally, DualNet-MP based on
the proposed SMAPE loss function 
achieves 4-dB performance
improvement in terms of NMSE reduction
for $\text{CR}_\text{PHA}$=1/8 at outdoor 
and 7-dB improvement for $\text{CR}_\text{PHA}$=1/8 
at outdoor. 

\subsection{Different Core Layer Designs}
To investigate the appropriate core layer designs of DualNet-MP
in order to efficiently extract the underlying features 
of CSI phases, we provide a performance evaluation using 
FC, 
linear convolutional, and circular convolutional layers, 
respectively, for the core network.
Denoted respectively as DNN, CNN and C-CNN,
these networks adopted SSQ \cite{CQNET} and binary-level quantization (BLQ) as the quantization module at
the encoder. 
Denote that the DNN design follows the recent work \cite{CoCsiNet}.
We consider the phase compression ratio of $\text{CR}_\text{PHA} = 1/8$. For SSQ, we assign $K_\text{PHA} = 8$ bits for each codeword. That is, there are  $\text{CR}_\text{PHA}Q_{t}N_b = 128$ 
8-bit codewords sent to the gNB. In contrast, there are  $K_\text{PHA}\text{CR}_\text{PHA}Q_{t}N_b = 1024$ 
1-bit codewords when applying BLQ.

\begin{figure}
\setlength{\abovecaptionskip}{0.cm}
\setlength{\belowcaptionskip}{-0.cm}
\centering
\resizebox{3.8in}{!}{
\includegraphics*{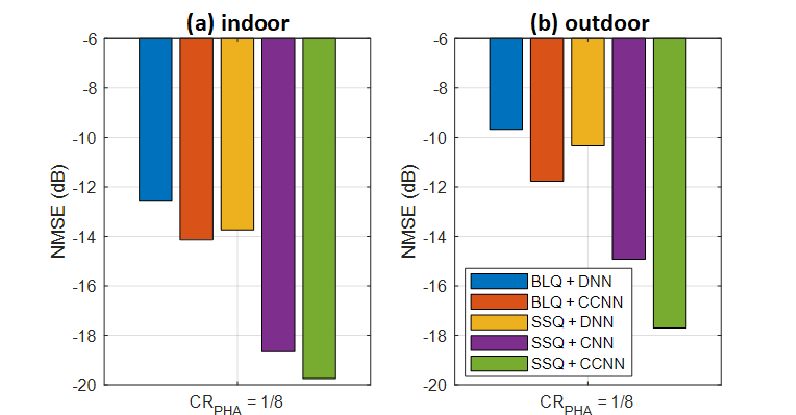}}
\caption{NMSE performance for different core layer designs in (a) indoor and (b) outdoor scenarios.\label{expC}}
\end{figure}

Figs. \ref{expC}.(a) and (b) show the NMSE performance for 
the considered core layer designs. 
For both indoor and outdoor scenarios, DualNet-MP 
demonstrates superiority when adopting SSQ and C-CNN,
which can be attributed to 
two possible reasons. Firstly, unlike BLQ, 
SSQ is differentiable such that it is easier to train. 
Secondly, there are many structural and circular features 
of CSI phases in the angle-delay domain 
that can be extracted better with the proposed structural
changes. 

{In terms of storage and complexity of the proposed architecture, 
we note that C-CNN with only 826K parameters
is considerably simpler
than DNN of 11.6M parameters, using
comparable floating point operations.} Thus, the newly proposed DualNet-MP architecture 
combining SSQ and C-CNN provides both performance gain
and cost benefits. 

\subsection{Accessment of FLOPs and parameter numbers}

We evaluated the numbers of floating point operations (FLOPs) and parameters of DualNet-MP with different phase encoding approaches and core layer designs in Tables \ref{Complexity_Diff_Encoding} and \ref{Complexity_Diff_Layer_Design}, respectively.
The comparison shows similarly 
complexity in terms of FLOPS. Since the complexity and storage issues are both crucial in the design of encoders
deployed at UEs, we provide Table \ref{Complexity_Compare} to compare the encoder complexity between the proposed framework and two state-of-the-art alternatives. 

\begin{table}
\centering
\caption{FLOP and parameter numbers of DualNet-MP with different phase encoding schemes.\label{Complexity_Diff_Encoding}}
\begin{tabular}{|l|c|c|c|c|}
\hline
           & \multicolumn{2}{c|}{MDPQ}     & \multicolumn{2}{c|}{$\text{MSE}_\text{0}$}   \\ \hline
$\text{CR}_\text{PHA}$         & \multicolumn{2}{c|}{1/8,1/16} & 1/8          & 1/16        \\ \hline
FLOPs      & \multicolumn{2}{c|}{39.5M}    & 77.2M        & 76.9M       \\ \hline
Parameters & \multicolumn{2}{c|}{282.4K}   & 825.9K       & 694.7K      \\ \hline
           & \multicolumn{2}{c|}{MDPP}     & \multicolumn{2}{c|}{SMAPE} \\ \hline
$\text{CR}_\text{PHA}$         & 1/8           & 1/16          & 1/8          & 1/16        \\ \hline
FLOPs      & 77.2M         & 79.6M         & 79.2M        & 78M         \\ \hline
Parameters & 825.9K        & 694.7K        & 826K         & 695.3K      \\ \hline
\end{tabular}
\end{table}

\begin{table}
\centering
\caption{FLOP and parameter numbers of DualNet-MP with different core layer designs when $\text{CR}_\text{PHA} = 1/8$.\label{Complexity_Diff_Layer_Design}}
\begin{tabular}{|l|c|c|c|c|c|}
\hline
                  & \multicolumn{2}{c|}{DNN} & \multicolumn{2}{c|}{C-CNN} & CNN  \\ \hline
Quantization Type & BLQ         & SSQ        & BLQ          & SSQ         & SSQ   \\ \hline
FLOPs             & 118M        & 81.7M       & 82.8M        & 79.2M       & 79.2M \\ \hline
Parameters        & 34.2M       & 11.8M      & 2.6M         & 826K        & 826K  \\ \hline
\end{tabular}
\end{table}

\begin{table}
\centering
\caption{FLOP and parameter numbers at the encoders of DualNet-MP and state-of-the-art alternatives when $\text{CR} = 1/8$.\label{Complexity_Compare}}
\begin{tabular}{|l|c|c|}
\hline
Method     & Parameters & FLOPs \\ \hline
CsiNet+    & 525K      & 1.86M   \\ \hline
CoCsiNet   & 14.9M     & 39.1M \\ \hline
DualNet-MP & 280K      & 38M   \\ \hline
\end{tabular}
\end{table}

Since the FC layers are responsible for
most of the parameters in many models,  CoCsiNet requires the most parameters as it uses multiple FC layers at the encoder. Compared with CsiNet+, the proposed framework requires about half storage by separately encoding CSI phase and magnitude. To be specific, the FC layers require
$(2048\times256 + 256)$ and $2(1024\times128 + 128)$ parameters at the encoders of CsiNet+ and DualNet-MP, respectively.

Often, convolutional layers contribute most FLOPs in a model due to the considerable number of two-dimensional convolution operations. However, DualNet-MP uses more
convolutional layers at the encoder 
but still requires fewer FLOPs than CoCsiNet. 
Thus the proposed new architecture exhibits
improvement in complexity reduction.

\section{Conclusions}
This work presents a new deep-learning (DL) framework 
for large scale CSI estimation 
that leverages feedback compression and auxiliary
CSI magnitude information in FDD systems.
{Utilizing vital domain knowledge in
DL for CSI estimation to
overcome known training issues, our new
framework provides a novel loss function to
enable efficient end-to-end learning and
improves CSI recovery performance.} 
We further exploit
the circular characteristics of the underlying
CSI in DA domain to propose an innovative
circular convolution neural network (C-CNN). 
Our test results reveal
significant improvement of
overall CSI recovery performance for both
indoor and outdoor scenarios 
and complexity reduction in comparison 
with a number of published 
alternative DL compression designs
for MIMO CSI feedback.

\bibliography{references.bib}
\bibliographystyle{IEEEtran}
\end{document}